\documentclass{article}
\usepackage{graphicx}

\title{Forbush Decrease Data with a Simple Model}

\author{C. D'Andrea, J. Poirier, D. S. Balsara \\
{\it Department of Physics, University of Notre Dame} \\ 
{\it  Notre Dame, Indiana, USA} 
}

\date{}

\begin{document}

\maketitle

\begin{abstract}

On October 28, 2003 an earthward-directed coronal mass ejection (CME) was observed from SOHO/LASCO 
imagery in conjunction with an X17 solar flare.  The CME, traveling at
nearly 2000~km/s, impacted the Earth on October 29, 2003 causing ground-based particle
detectors to register a counting rate drop known as a Forbush decrease.  In
addition to affecting the rate of cosmic rays, the CME was also responsible for causing
anisotropies in the direction of incidence.  Data from Project GRAND, an
array of proportional wire chambers, are presented during the time of this Forbush decrease. 
A simple model for CME propagation is proposed and we present an argument based on gyroradius
that shows that a magnetic field of the radius calculated for the ejecta is sufficient to
deflect energetic charged particles of an energy detectable by GRAND.

\end{abstract}

\section{Introduction}

In late October of 2003 the Earth experienced an extraordinary amount of solar and
geomagnetic activity originating from solar region NOAA 10486, including an X17 flare
which was one of the largest solar flares since 1976 \cite{IPS03}.  This flare was
detected beginning at 9:51~UT on October 28, 2003 (day 301) and peaked at 11:10~UT
the same day.  This flare had an associated CME with a transit time from the sun
to the Earth of only 19 hours, making it one of the the fastest on record.  The shock
from the CME impacted the Earth's magnetic field as a strong sudden impulse (an
abrupt increase in the horizontal component of the geomagnetic field) at 6:13 UT on
October 29, 2003 (day 301) \cite{NOAA}.  This CME caused a drop in the counting rate
of ground-based cosmic ray detectors.  The counting rate for neutron
monitor stations remained suppressed for roughly twelve days after the impact of the
shock.

Project GRAND \cite{Poirieret03}, an extensive air shower array of
proportional wire
chambers, is a useful tool for studying the ground-level effects of
cosmic rays.  GRAND's median energy for vertically incident
cosmic rays is 56 GeV, higher than that for neutron monitors.
This allows GRAND to complement neutron monitor
data by studying higher energy effects.  GRAND has an angular
resolution of $0.26^{\circ}$ on a projected plane for incoming muon tracks.  
The angular resolution, primary energy sensitivity, and large detector area of
this experiment (82~$m^2$) make it an excellent instrument to
to study Forbush decreases.
Secondary muon data obtained by GRAND during the time of the October 29
Forbush decrease are discussed.  In addition, a simple model for a CME ejecta 
is proposed and some of the consequences of this model are discussed.  

\section{Project GRAND}

Project GRAND is located at $41.7^\circ$ N and $86.2^\circ$ W at 220 meters
above sea level.  Its array of 64 proportional wire chamber stations are
arranged in a 8 $\times$ 8 grid that covers 100~m $\times$ 100~m.  Each station
contains four pairs of wire chamber planes.
Each pair has a top plane with wires running north / south and a bottom plane
with wires running east / west.  Each plane contains 80 wire cells with a
total active area of 1.29~$m^2$.  The pairs of planes are placed
vertically above one another, with a separation of 197~mm between pairs.
There is a 51~mm steel plate above the bottom pair of planes, allowing muons to be
differentiated from electrons (which scatter, shower, or
stop in the steel).  The angle of the incoming tracks is determined from
the difference between the location of the wires hit
on the
top plane and the bottom plane.  The stations have a maximum sensitivity
for vertical tracks and a cutoff of $63^{\circ}$ from vertical
in each projected plane due to the size and separation of the planes.  GRAND runs with two
triggers: 1) single tracks from individual stations and 2) multiple stations in time
coincidence.  The data presented in this paper comes solely from trigger (1) analyzed for
those single tracks which are identified as muons.

\begin{figure}
\begin{center}
\includegraphics[width=20pc]{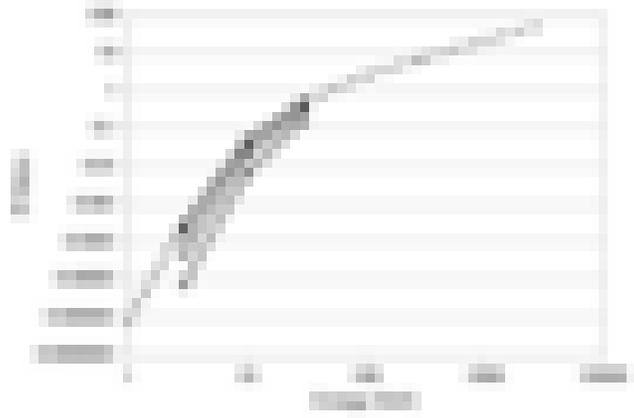}
\caption{
Results from primary protons simulated by the FLUKA Monte Carlo code. The
R-value is the number of muons reaching ground level, on average, from a primary proton
with the given energy. The dashed
line represents vertically incident particles while the solid lines below the dashed
represent (from top to bottom) $13^\circ$, $26^\circ$, and $39^\circ$ inclination from 
vertical.
}
\label{FLUKAFig}
\end{center}
\end{figure}

The Monte-Carlo program FLUKA \cite{FLUKA},\cite{FLUKA2} was used to simulate
primary protons in the atmosphere for energies of interest (1~-~
3000~GeV).  The results of these simulations were originally shown in
\cite{JGR} for primary protons and
\cite{Poirieret02} for primary gamma rays.  The results of the number
of muons reaching ground level per proton of a given primary energy are shown in Figure
\ref{FLUKAFig}.  Protons were used for this simulation because the
majority of the primaries which generate ground level muon counting rates
are protons

\begin{figure}
\begin{center}
\includegraphics[width=20pc]{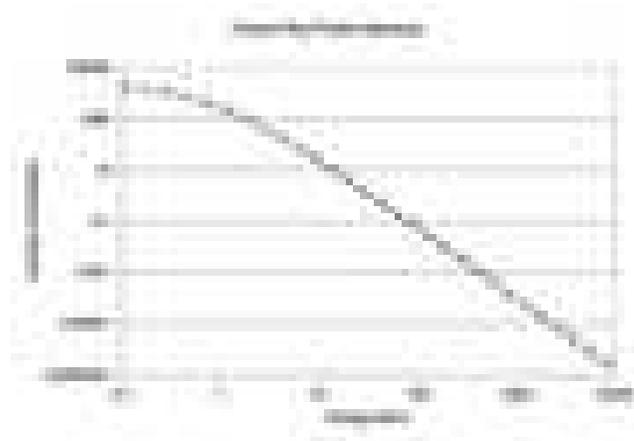}
\caption{
The graphical representation of the primary proton spectrum of
$N(E)= 1.74\times10^4(E+0.89)^{-2.75} dE$
given by \cite{Fujimoto}.
}
\label{Primary}
\end{center}
\end{figure}

\begin{figure}
\begin{center}
\includegraphics[width=20pc]{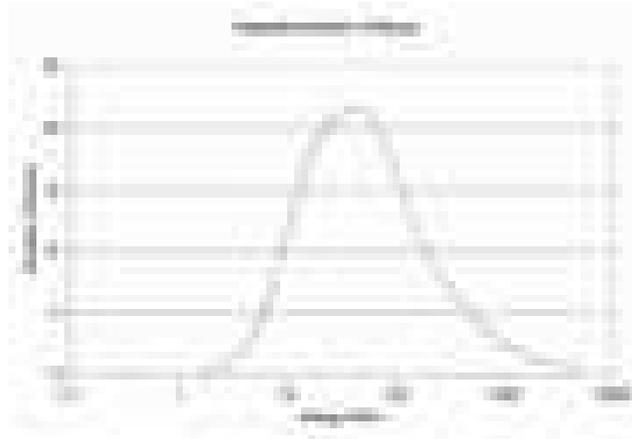}
\caption{
GRAND's sensitivity to primary protons of various energies.  GRAND is most sensitive to
vertical protons with an energy of 56 GeV.
}
\label{Response}
\end{center}
\end{figure}

The response of GRAND to background cosmic rays can be determined by using
the results from the FLUKA simulation and folding it with the cosmic ray
spectrum at those angles.  A primary spectrum of
\begin{center}
\begin{equation}
N(E)= 1.74\times10^4 (E+0.89)^{-2.75} dE
\end{equation}
\end{center}
\noindent
was used \cite{Fujimoto} and is shown in Figure \ref{Primary}.
The combined response to
vertical primary protons is shown in Figure \ref{Response}, showing
GRAND's median primary rigidity at 56 GV.

Further information on the response function and operation of Project
GRAND is available from
\cite{JGR}, \cite{Poirieret03}, and references therein.

\section{Data}

\begin{figure}
\begin{center}
\includegraphics[width=20pc]{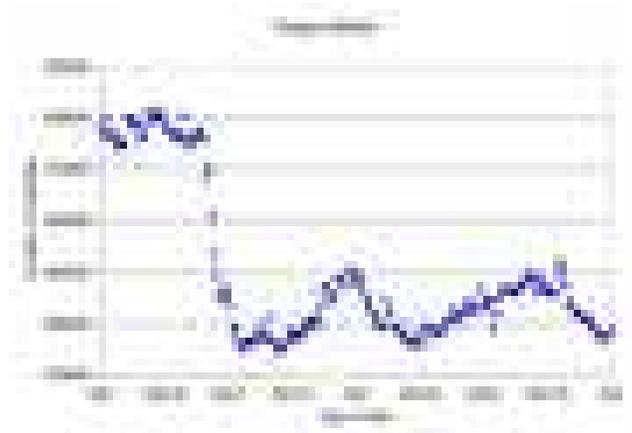}
\caption{Data from Project GRAND and Nagoya for October 28, 2003 (D301) and the following
day.  The data show the counting rate for each 15 minute bin.  There is an 8\% decrease
due to the Forbush decrease.}
\label{Grand}
\end{center}
\end{figure}

The muon data rate for October 29 and 30, 2003 is shown in Figure
\ref{Grand}.  GRAND's data rate shows a decrease of 8\% from the
counting rate prior to the decrease.  Data following the Forbush decrease show
that it takes GRAND nine days to recover to its original counting rate.

In order to ensure that the data reflected only physical variations, a cut was performed
to select only the best stations for analysis for this event.  The r.m.s. deviation
was calculated for each of the stations and compared to the expected
statistical fluctuation
(the square root of the mean number of counts).  All stations with a ratio higher
than 8.0 were eliminated from this analysis, leaving 17 stations.

\begin{figure}
\begin{center}
\includegraphics[width=20pc]{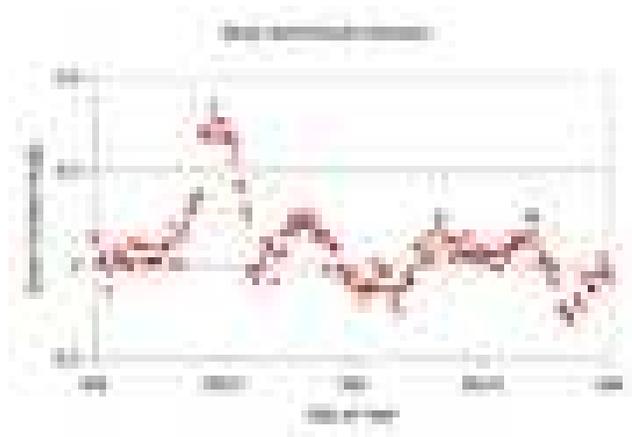}
\caption{GRAND's
angular data expressed as the mean angle (in degrees) in the north/south direction.  Note the
variation in the north/south direction during the time of the Forbush decrease.}
\label{Angle}
\end{center}
\end{figure}

Angular information for the muons was also analyzed during October 29 and October 30.
The average angles in the north/south direction and the
east/west direction were calculated and are shown in Figure
\ref{Angle}.  While the east/west direction shows
little change in activity above statistics, the north/south angle
shows a swing where more particles originate from the north during the time of the Forbush
decrease.  Since a Forbush decrease is caused by a large solar coronal
ejection whose magnetic field deflects the incoming cosmic ray particles
when it impacts the Earth, it should be expected that this
causes a deficiency in particles from a particular direction.  

\begin{figure}
\begin{center}
\includegraphics[width=20pc]{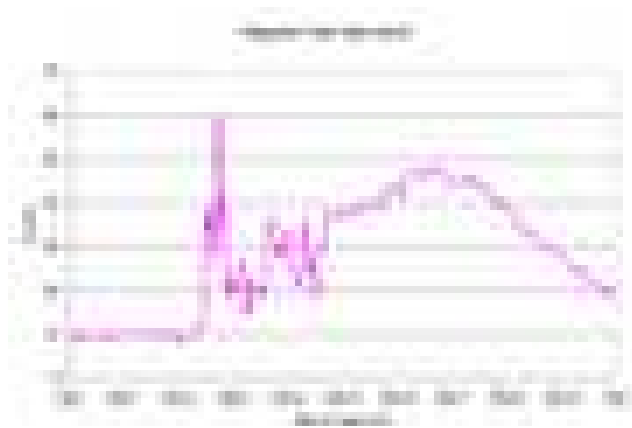}
\caption{Magnetic field data from ACE for October 29, 2003 (Day 302).  Note the
quiet magnetic field (B $\le$ 10~nT (1 nanotesla = 10$^{-5}$ Gauss) before the arrival of the 
CME and the sharp increase at the time of arrival.}  
\label{ForbushACE}
\end{center}
\end{figure}

The magnetic field near the Earth was also studied \cite{ACE} during the
time of
this event and is shown in Figure \ref{ForbushACE}.  The magnetic field shows an
increase during the time
of the decrease which maintained for a day following the impact.  This
is consistent with a large coronal mass ejection with its own magnetic
field interacting with the Earth's magnetic field at the time of impact.
This increased magnetic field can result in a higher cutoff rigidity for
cosmic rays near the Earth, resulting in a decrease in the counting rate
as seen from the ground.

\section{Theoretical Model}

Decreases in cosmic-ray intensity in coincidence with changes in the
magnitude of the horizontal component of the geomagnetic field
(the component of the magnetic field parallel to the Earth) were
first noticed by Scott Forbush in the 1930s.
\cite{Morrison} stated that the sun would occasionally emit magnetized plasma from
active regions.  These clouds would affect the cosmic-ray rate in
interplanetary space and produce magnetic storms on the Earth.  
Satellites were
able to detect changes in the rate of cosmic rays (along with a
simultaneous increase in magnetic field strength) in interplanetary
space in 1959 and 1960 confirming this hypothesis \cite{NAS57}.
These magnetic clouds have been identified as the
ejecta from coronal
mass ejections on the sun.  In order to properly understand the mechanisms of
Forbush decreases it is necessary to study the propagation of ejecta in the solar
wind \cite{Cane2000}.

CMEs often originate in the solar corona near
magnetic field lines and typically follow a coronal helmet streamer
\cite{GibsonLowe}.
This streamer gets distorted and is finally disrupted by the expanding closed
field region underneath it. The CME speed is
typically between 20 and 2000 km/s and have an average speed of 400 km/s
\cite{Cane2000} which is greater than the typical solar wind speed from
the sun.  When
the fast ejecta meets with the slower solar wind, a shock is
created \cite{Sokolov}.  It should be noted that while
CMEs and flares often are
associated, one is not necessary for the other.  

A simple model for the propagation of the ejecta generated by a CME is presented
here.  With a given
initial radius and magnetic field near the
surface of the sun, the intended goals are to determine how it propagates through the
solar system, the time of flight from the sun to the
Earth, the magnetic field of the ejecta when it reaches the
Earth, and the extent of the ejecta at Earth.  The information on the size and
magnetic field  will be used to evaluate the prospect that high energy cosmic rays may be
occluded by the material.  The model presented here is a simple model that
assumes a bulk size, shape, pressure, and magnetic field.  This may not be the
case if the structure created during the formation of the CME is maintained.  The
model, however, is adequate for determining scale values for the magnetic field
and size of the ejecta when it impacts the Earth.

The first assumption is that it is the solar wind which is responsible for convecting the 
CME material to the Earth and therefore the flight time for the CME ejecta is connected to 
the solar wind.  Any relative motion between the CME ejecta and
the solar wind would dissipate as a shock or pressure fluctuation.  The time
scale for the solar wind to travel from the sun to the Earth and the time scale
of a CME ejecta transit are quite similar, positing a relationship between the
two which is explored further in Section 5.

\begin{figure}
\begin{center}
\includegraphics[width=20pc]{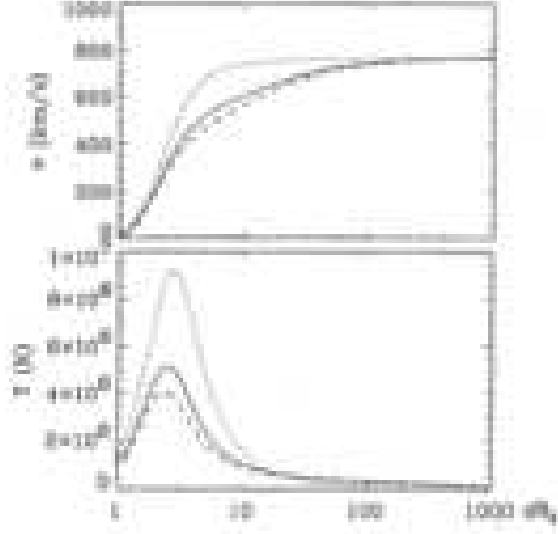}
\caption{The solar wind velocity and
temperature versus distance from the sun.  This shows the effect of a fraction of undamped waves on the fast 
solar wind model. Dotted line (all waves damped), full line (5\% undamped), dashed line (10\% 
undamped).This 
figure was originally printed in \cite{Axford} as Figure 4 and is used with permission. 
Only the dotted lines 
(all waves damped) were used in this work.}
\label{axpic}
\end{center}
\end{figure}

In order to construct a simple model, it is
necessary to know the properties of the solar wind as it propagates from
the sun to the Earth.  Pressure balance between the solar wind and the CME ejecta is
then used to determine the properties of the CME ejecta as they flow toward the Earth.
The results of a calculation of solar wind
velocity and temperature with respect to distance from the sun are given
in \cite{Axford}, shown here as  Figure \ref{axpic}.  For
the purposes of discussion in this work only the damped waves will be
considered (dotted lines). These values of $v_w$ and $T_w$ from Figure \ref{axpic}
are then used to determine the density of the
solar wind at a given distance from the sun using a few basic
principles.

We assume that the mass of the wind is constant as it expands and that its
expansion is spherically symmetric:
\begin{center}
\begin{equation}
\label{mass_cons}
\rho_w v_w d^2 = constant
\end{equation}
\end{center}
\noindent
where $d$ is the distance from the sun, $v_w$ is the velocity of
the solar
wind, and $\rho_w$ is the density of the solar wind.  Also it is
assumed
that the wind and CME follow the ideal gas law. The mass continuity equation is used
to derive an equation for the density based on initial conditions:
\begin{center}
\begin{equation}
\rho_w v_w d^2 = \rho_{w0} v_{w0} d_0^2
\end{equation}
\begin{equation}
\rho_w = \frac{\rho_{w0} v_{w0}}{v_w}(\frac{d_0}{d})^2
\label{densityEQ}
\end{equation}
\end{center}
\noindent
This equation for density is
then used to determine the density of the wind at all distances given the
velocity of the solar wind and the distance from the sun. \cite{Axford}
gives $\rho_0$ of $10^8$ particles/$cm^3$ and a $v_{w0}$ of 3.3$\times 10^6$~cm/s at a
$d_0$=6.96$\times10^{11}$~cm (one solar radius).

The pressure of the solar wind will be important in determining the size
of the CME as it propagates through the solar system.  The total pressure of the
CME is balanced by the pressure of
the solar wind.  As the
pressure due to the solar wind decreases, the CME expands.  In this
case we assume that the solar wind behaves like an ideal gas.  We use $\mu$ to
represent the effective mass.  From the ideal gas law, the gas pressure of the
solar wind is $P_w = R \rho_w T_w / \mu$.  There is also a contribution to
the pressure in the solar wind from
the interplanetary magnetic field, making the total pressure due to the solar wind (in cgs 
units):

\begin{center}
\begin{equation}
P_w = \frac{R \rho_w T_w}{\mu} + \frac{B_w^2}{8 \pi}
\end{equation}
\end{center}
\noindent
The values for $\rho_w$ can be found in Equation (\ref{densityEQ}) and
the values for $T_w$
are those from \cite{Axford}.  The values for $B_w$ assume a
Parker spiral \cite{Parker} which has the
form of $B_w = B_{w0}(d_0/d)^2$.  A value of 7~nT is
used for $B_{w0}$ at a distance $d_0 = 1~AU$.  This value is a mean value
of the
magnitude
of the IMF as recorded by ACE on October 27, 2003 \cite{ACE}.
The initial density of the solar wind at the sun was given in \cite{Axford}.
Since $\rho_w$ can be found at all distances,
$d$, and since $T_w$ can be read from Figure \ref{axpic}, $P_w$ can be determined
at all distances, $d$.

In order to maintain dynamical equilibrium at each radius, the pressure of the CME
ejecta continually adjusts to match that of the ambient solar wind.  The
CME ejecta
has contributions to the pressure from magnetic pressure and gas pressure.  The
adiabatic equation is assumed and gives the gas pressure of the CME, $P_c$:

\begin{center}
\begin{equation}
P_c=P_{c0} (\frac{\rho_c}{\rho_{c0}} )^{\gamma}
\end{equation}
\end{center}

The development of $\rho_c$ is determined by the geometry of the CME and is
dependent on its formation in the solar corona and the structure of the magnetic
field within it.  Typically the CME structure can be complex, for example
the
structure given in \cite{GibsonLowe} is a CME with a three-part structure, the center of
which is typically an erupted solar prominence.  This is surrounded by a
cavity which is, in turn, surrounded by a leading plasma loop.  This gives a CME
structure which is a magnetic field embedding a magnetic flux rope which originates
at the base near the corona.  If the height of the eruption is similar to the width
of the loop of the magnetic field lines, the CME will have a mostly spherical
structure.  If the height of the prominence is larger than the radius of
the magnetic field loop then the structure of the CME will be more cylindrical
in nature.  Both geometries are explored below.

For a CME with spherical geometry, the continuity equation is
$\rho_c r_c^3 = constant$.  For a CME with cylindrical geometry, the
continuity
equation changes to $\rho_c r_c^2 = constant$, where $r_c$ is the radius of the CME
ejecta.  We are assuming here that cylindrical CME ejecta might arise from
an arcade flare-like structure.  In general, the spherical and cylindrical
geometries represent two structural extremes and we can also envision an
intermediate case with $\rho_c r_o^{\beta} = constant$
with $2 \leq \beta \leq 3$.  For
this
work we used $\gamma\beta=4.0$.We explore the spherical (denoted Sph.),
cylindrical (denoted
Cyl.), and intermediate (denoted Int.) cases below.  Substituting these relationships
into the pressure equation above, we get the development of the gas
pressure of the CME with respect to radius.

\begin{center}
\begin{eqnarray}
\hspace{1.0in} P_c &=& P_{c0} (\frac{r_{c0}}{r_c})^{3\gamma} \; \; Sph. \nonumber\\
\hspace{1.0in} P_c &=& P_{c0} (\frac{r_{c0}}{r_c})^{\beta \gamma} \; \; Int.  \\
\hspace{1.0in} P_c &=& P_{c0} (\frac{r_{c0}}{r_c})^{2\gamma} \; \; Cyl. \nonumber
\end {eqnarray}
\end{center}

\noindent
Conservation of magnetic flux in the CME ejecta requires that
\begin{center}
\begin{equation}
B_c r_c^2 = B_{c0} r_0^2.
\label{Mag}
\end{equation}
\end{center}
\noindent
Combining that with the magnetic pressure and substituting this
into the magnetic
pressure equation gives,
\begin{center}
\begin{equation}
P_{Bc} = \frac{B_{c0}^2}{8\pi}(\frac{r_{c0}}{r_c})^4
\end{equation}
\end{center}
\noindent
As a result, assuming that the CME ejecta are in pressure balance with the ambient solar
wind as they propagate toward Earth, we get:

\begin{center}
\begin{eqnarray}
\frac{R \rho_w T_w}{\mu} + \frac{B_w^2}{8 \pi} &=&
P_{c0} (\frac{r_{c0}}{r_c})^{3\gamma} +
\frac{B_{c0}^2}{8\pi}(\frac{r_{c0}}{r_c})^4 \; \; Sph. \nonumber \\
\frac{R \rho_w T_w}{\mu} + \frac{B_w^2}{8 \pi} &=&
P_{c0} (\frac{r_{c0}}{r_c})^{\beta \gamma} +
\frac{B_{c0}^2}{8\pi}(\frac{r_{c0}}{r_c})^4 \; \; Int.  \\
\frac{R \rho_w T_w}{\mu} + \frac{B_w^2}{8 \pi} &=&
P_{c0} (\frac{r_{c0}}{r_c})^{2\gamma} +
\frac{B_{c0}^2}{8\pi}(\frac{r_{c0}}{r_c})^4 \; \; Cyl. \nonumber
\end{eqnarray}
\label{CompleteEqns}
\end{center}
\noindent
The initial radius is based on the final reported radius
for the October 28, 2003 CME 
\cite{SOHO} which was $r_{c0}=4.49\times10^{11}~cm$.
The initial magnetic field of 1.0~G is a typical field at that distance from
the sun and was chosen based on estimates made in
\cite{Cane2000}.
The results were also
determined for magnetic fields of 0.3 and 3.0 Gauss.  This range of an order
of magnitude in initial magnetic field gives information on initial magnetic
pressure for the CME over the range of two orders of magnitude.  $P_{c0}$ was
obtained by solving each of the equations shown in
Equation (\ref{CompleteEqns}) using
initial conditions and $r_c = r_{c0}$.  Finally, $r_c$, the
size of the CME ejecta,  was
solved numerically for different values of solar wind pressure and temperature
(effectively solving for $r_c$ as a function of distance from the sun).  Once that
is done, Equation (\ref{Mag}) can be used to solve for the magnetic field as
a function of radius.

\section{Comparison with Data}

The flight time of the CME to reach one AU was calculated and is independent of
the geometry. The flight time predicted by the model was 54 hours.  The typical 
values for CME speeds range from 20 - 2000 km/s \cite{Cane2000} which correspond 
to flight times from 21 hours to 86 days.  The predicted flight time falls within 
this range.  The measured time of flight was 19 hours.

\begin{figure}
\begin{center}
\includegraphics[width=20pc]{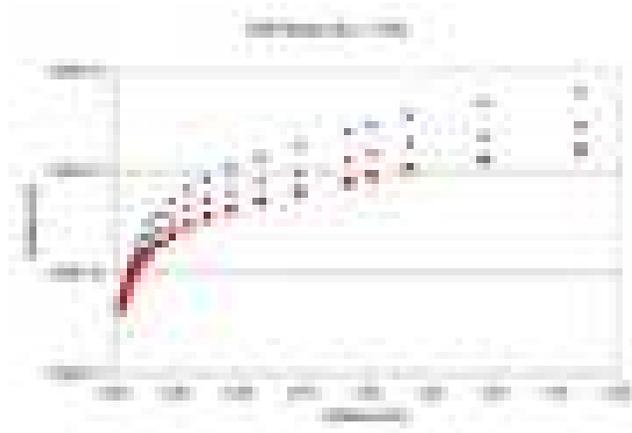}
\caption{The radius of the CME at various distances from the sun as predicted
by the CME models in this paper.  Circles represent the spherical model, squares
represent the cylindrical model and diamonds repreent the intermediate model.  All the
models assumed an initial CME magnetic field of 1~G (1 nanotesla = 10$^{-5}$ Gauss).}
\label{modelR}
\end{center}
\end{figure}

\begin{figure}
\begin{center}
\includegraphics[width=20pc]{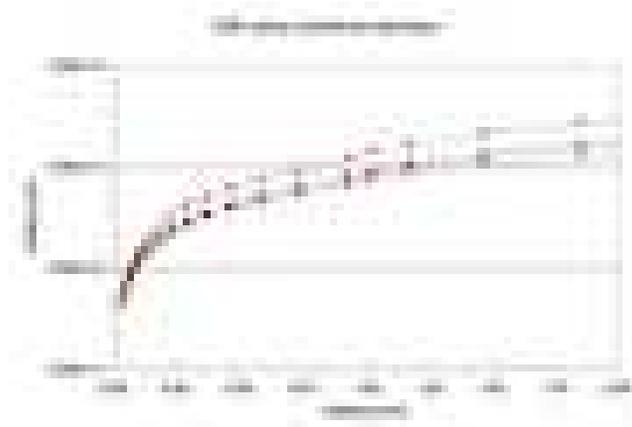}
\caption{The progression of the radius of the CME using the cylindrical model for three
different values for its inital magnetic field. Solid lines denote an initial value for
the
magnetic field of 1~G while the dashed lines below and above the solid lines denote
initial conditions of 0.3~G and 3~G, respectively.}
\label{CylEx}
\end{center}
\end{figure}

The radial extent of the CME at 1 AU was determined for all three geometrical cases and
for a range of magnetic fields.  If an initial magnetic field of 1.0~G is used, the
spherical situation gives a CME with a radius of
1.9~AU and the cylindrical CME has a radius of
0.60~AU and the intermediate geometry CME has a radius of
0.98~AU.  The speed of the solar wind near the Earth as
reported by ACE \cite{ACE} during the event (5:00 - 19:00~UT) was used to determine
the
size of the CME.  This gives a CME size of 0.54~AU.
The progression of the radial size as a function of
distance from the sun for all three geometries is shown in Figure
\ref{modelR}.
Figure \ref{CylEx} shows the radius as a function of
distance for the cylindrical geometry for all three inital magnetic field strengths.

\begin{figure}
\begin{center}
\includegraphics[width=20pc]{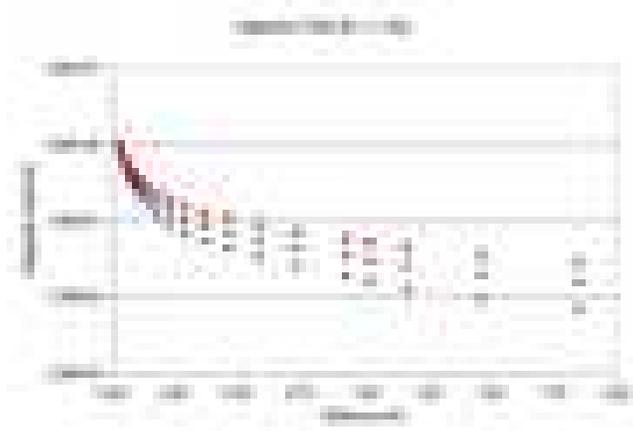}
\caption{The magnetic field of the CME at various distances from the sun as
predicted given by the CME models in this paper.  Circles represent the spherical
model, squares
represent the cylindrical model and diamonds represent the intermediate model.  All
models in this figure used a 1~G initial magnetic field for the CME.}
\label{ModelB}
\end{center}
\end{figure}

\begin{figure}
\begin{center}
\includegraphics[width=20pc]{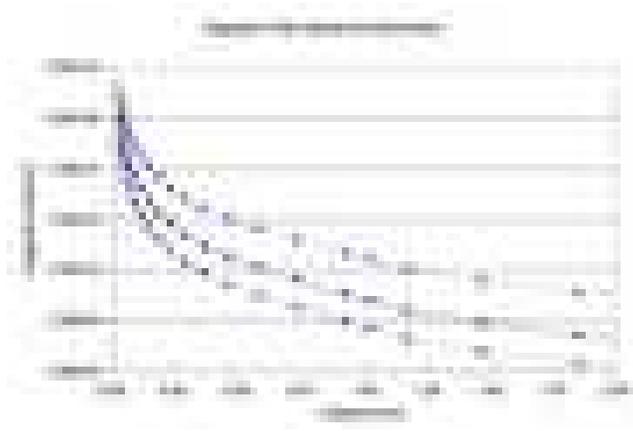}
\caption{The progression of the magnetic field of the CME as a function of distance
from the sun using the spherical model and three different inital magnetic field values.
Solid lines denote an
initial value for the
magnetic field of 1~G while the dashed lines above and below the solid lines denote
initial conditions of 0.3~G and 3~G, respectively.}
\label{SphEx}
\end{center}
\end{figure}

The magnetic field of the CME was also calculated at various points between the sun
and the Earth.  These are shown in Figure \ref{ModelB}.  Given an initial magnetic 
field of 1~G, the cylindrical case gives a mean magnetic
field of 250~nT at 1 AU while the spherical case shows a mean field of 24~nT and
the intermediate case gives a mean field of 90~nT.  The range in magnetic field for
the CME at 1 AU is between 6.9 and 160~nT for the spherical case while the
cylindrical model yields a range between 110 and 310~nT and the intermediate model
gives a range from 28 to 280~nT.  Data from ACE show that the IMF fluctuates
during this storm between 20 and 60~nT \cite{ACE}.  Figure \ref{SphEx} shows the 
progression
of the magnetic field for the spherical model given the three different initial
magnetic fields.  A summary of 
the results for all three models and all initial magnetic fields is shown in Table 
\ref{Table}.

The three models presented here provide a parameter range that agrees
with the data for this event.  Given the simplicity of this model, an exact fit was
not expected.  The spherical model gives a value for the
magnetic field of the CME at Earth that is consistent with that observed during the
Forbush decrease of October 29, 2003 where  The cylindrical model gives a
value for the CME radius more consistent with the data.
It is also interesting to note that the CME observed is from one
of the largest flares on record and is most likely different from a ``typical"
CME.  This model is adequate if used as a scaling argument.

If it is assumed that GRAND's drop of 8\% in its counting rate is caused by an
increased magnetic field which prohibits lower energy primaries from
reaching the Earth, the minimum energy primaries observed by GRAND during the
decrease
can be determined.  Eliminating the lower 8\% of GRAND's cosmic rays leaves
primaries with energies above 10~GeV.  Given $pc=Br$, where $pc$ is
effectively the energy of
a charged particle, B is the magnetic field, and $r$  is the gyroradius for a
particle in that field, a 10~GeV particle in a 24~nT field has a gyroradius of
0.0087~AU, larger than the Earth's radius and much smaller than the size
of the CME.  This implies that a magnetic field of the size and strength calculated
is strong enough to have an effect on GRAND's counting rate.

\section{Conclusions}

Project GRAND sees an 8\% drop in its secondary muon counting rate during the Forbush
decrease of October 29, 2003.  A shift is also observed in the mean angle of incident
muons in the north-south plane.  A simple model for the propagation of a CME
through the solar system is presented.  Within a range of initial conditions for
the CME, the model is  consistent with satellite
observations near Earth during the October 29, 2003 storm.

\section{Acknowledgments}
We thank Mitch Wayne and quarknet, Michael Wiescher
and JINA, and Terry Rettig for funding and support.  We thank Nathan
Johnson-McDaniel for his
assistance.  D.~S.~Balsara acknowledges support via NSF grants R36643-7390002,
AST-005569-001, AST-0607731 and NSF-PFC grant PHY02-16783.  Project
GRAND was constructed through grants from the National Science
Foundation and is funded through the University of Notre Dame and private donations.

\begin {table}
\begin{center}
\begin{tabular} {|c|c|c|} \hline \hline
Model & CME Radius, $r_c$ & CME B Field, $B_c$ \\
\hline \hline
$S_{0.3}$: Sph.; $B_{c0}$=0.3~G & $1.7~AU$ & $6.9~nT$ \\
$S_{1}$: Sph.; $B_{c0}$=1~G & $1.9~AU$ & $24~nT$ \\
$S_{3}$: Sph.; $B_{c0}$=3~G & $2.0~AU$ & $160~nT$ \\
\hline
$I_{0.3}$: Int.; $B_{c0}$=0.3~G & $0.98~AU$ & $28~nT$ \\
$I_{1}$: Int.; $B_{c0}$=1~G & $0.98~AU$ & $90~nT$ \\
$I_{3}$: Int.; $B_{c0}$=3~G & $0.98~AU$ & $280~nT$ \\
\hline
$C_{0.3}$: Cyl.; $B_{c0}$=0.3~G & $0.50~AU$ & $110~nT$ \\
$C_{1}$: Cyl.; $B_{c0}$=1~G & $0.59~AU$ & $250~nT$ \\
$C_{3}$: Cyl.; $B_{c0}$=3~G & $0.94~AU$ & $310~nT$ \\
\hline
\hline
\end{tabular}
\caption{This table gives the values for CME radius and magnetic field 
for each of the three models with each of the three initial magnetic 
fields used in this paper.}
\label {Table}
\end{center}
\end{table}

\end{document}